\documentclass[12pt]{article}
\usepackage{graphicx}

\def\mydate{January 6, 2005\\January 14, 2011}

\newcommand{\gam}{\gamma}
\newcommand{\del}{\delta}
\newcommand{\tr}{{\rm Tr}}
\newcommand{\eps}{\epsilon}
\newcommand{\hp}{\hat{\phi}}

\newcommand{\be}{\begin{equation}}
\newcommand{\ee}{\end{equation}}
\newcommand{\bea}{\begin{eqnarray}}
\newcommand{\eea}{\end{eqnarray}}

\newcounter{sxn}

\newcounter{axn}

\date{}

\def\ignore#1{{}}

\newcommand{\beeq}{\begin{equation}}
\newcommand{\eneq}{\end{equation}}
\newcommand{\beqn}{\begin{eqnarray}}
\newcommand{\eeqn}{\end{eqnarray}}

\def\mybig{\displaystyle \strut }

\def\la{\raise.16ex\hbox{$\langle$}\lower.16ex\hbox{}  }
\def\ra{\, \raise.16ex\hbox{$\rangle$}\lower.16ex\hbox{} }
\def\go{\rightarrow}

\def\onehalf{ \hbox{${1\over 2}$} }
\def\onethird{ \hbox{${1\over 3}$} }
\def\Tr{{\rm Tr \,}}
\def\hA{{\hat A}}

\def\myfrac#1#2{{\mybig #1\over \mybig #2}}

\begin{document}

\thispagestyle{empty}

\baselineskip=12pt

{\small \noindent \mydate    \hfill OU-HET 478/2004}

{\small \hfill  hep-th/0408068}

\baselineskip=40pt plus 1pt minus 1pt

\vskip 3.5cm

\begin{center}

{\Large \bf Generalized Monopoles}\\
{\Large \bf in Six-dimensional Non-Abelian Gauge Theory}\\

\vspace{3.cm}
\baselineskip=20pt plus 1pt minus 1pt

{\bf  Hironobu Kihara,$^1$ Yutaka Hosotani$^1$ and Muneto Nitta$^2$}\\
\vspace{.3cm}
{\small {}$^1$\it Department of Physics, Osaka University,
Toyonaka, Osaka 560-0043, Japan}\\
{\small {}$^2$\it Department of Physics, Tokyo Institute of Technology,
Tokyo 152-8551, Japan}\\
\end{center}

\vskip 3.cm
\baselineskip=20pt plus 1pt minus 1pt

\begin{abstract}
A spherically symmetric monopole
solution is found in  $SO(5)$ gauge theory with Higgs
scalar fields in the vector representation in six-dimensional Minkowski spacetime. The action of the Yang-Mills fields is
quartic in field strengths.  The solution saturates the 
Bogomolny bound and is stable.  
\end{abstract}
\newpage

\clearpage


Long time ago Dirac showed that quantum mechanics admits
a magnetic monopole of quantized magnetic charge despite 
the presence of a singular Dirac string.\cite{Dirac:um, Wu:es}  
A quantized Dirac string is  unphysical entity in the sense 
that it yields no physical, observable effect.  
Much later 't Hooft and Polyakov showed that such 
magnetic monopoles emerge as regular configurations
in $SO(3)$ gauge theory with spontaneous symmetry breaking triggered
by triplet Higgs scalar fields.\cite{'tHooft:1974qc}-\cite{Prasad:kr}
't Hooft-Polyakov monopoles emerge in grand unified theory of 
electromagnetic, weak, and strong interactions as well.
Although a monopole has not been found experimentally as a single
particle, the existence of such objects has far reaching consequences.
In the early universe, monopoles might have beed copiously produced,
significantly affecting the history of the universe since then.
In strong intercations, monopole configurations are believed vital
for color and quark confinement.

In the superstring theory all matter and interactions including 
gravity are truely unified in ten spacetime dimensions.  Six
extra dimensions may be compactified in a small size, or the observed
four-dimensional spacetime can be a brane immersed in ten dimensional 
spacetime.  It is  important in this scenario  to
explore solitonic objects in higher dimensional spacetime,
which may play an important role in compactfying extra
dimensions, or in producing and stabilizing brane structures.
Recent extensive study of domain walls in supersymmetric theories,
for instance, may have a direct link to the brane world scenario.\cite{Nitta}
In this paper we explore and establish solitons with finite energies 
in higher dimensional spacetime.

The energy of 't Hooft-Polyakov monopoles is bound from below 
by a topological charge.   Monopole solutions  saturate such 
bound, thereby the stability of the solutions being guaranteed 
by topology.\cite{Bogomolny:1975de}
This observation prompts a question if there can be
a monopole solution in higher dimensions.
Kalb and Ramond introduced Abelian tensor gauge fields coupled to 
closed strings.\cite{Kalb}  Nepomechie showed  
that a new type of monopole
 solutions appear in those Kalb-Ramond antisymmetric tensor gauge 
fields.\cite{Nepomechie}
Their implications to the confinement\cite{Orland:1981ku}
and to ten-dimensional Weyl invariant spacetime\cite{Hosotani1} has been 
explored.
Topological defects in six dimensional Minkowski space-time as  
generalization  of Dirac's monopoles were also found.\cite{Yang:1977qv}
Tchrakian has investigated monopoles in non-Abelian gauge theory
in higher dimensions  whose action involves polynomials 
of field strengths of high 
degrees.\cite{Tchrakian:1978sf, Tchrakian2}
Further, it has been known that magnetic monopoles appear in the 
 matrix model  in the gauge connections describing  Berry's 
 phases on fermi states.
In particular, in the USp matrix model  they are  described by 
$SU(2)$-valued anti-self-dual connections.\cite{Itoyama}

The purpose of this paper is to present regular monopole 
configurations with 
saturated Bogomolny bound in $SO(5)$ gauge theory in six dimensions.
Although the existence of such solutions has been suspected by
Tchrakian for a long time, the explicit construction of solutions
has not been given.  We stress that the monopole solution 
presented below is the first example of a soliton in non-Abelian gauge theory in higher dimensions which is 
regular everywhere and has a finte energy. 

Let us recall that in 't Hooft-Polyakov monopoles in four 
dimensions, both $SO(3)$ gauge fields and scalar fields are in 
the vector representation.  In three space dimensions
the Bogomolny equations for those fields  match both in space indices 
and internal $SO(3)$ indices.  This correspondence seemingly 
becomes obscure when space dimensions are greater than three.  
A key to find correct Bogomolny equations is facilitated with 
the use of the Dirac or Clifford algebra.

Consider  $SO(5)$ gauge theory in six dimensions.  Gauge fields   
$A_{\mu}^{ab} = - A_{\mu}^{ba}$
are in the adjoint representation, whereas scalar fields $\phi^{a}$
are in the vector representation ($a,b= 1 \sim 5$).  To interrelate these two,
we introduce a basis $\{ \gamma_a \}$ of the Clifford algebra;
$\{\gamma_a , \gamma_b\} = 2 \delta_{ab}$ ($a,b = 1 \sim 5$).
We write $\phi \equiv \phi^a \gamma_a$ and $A = 1/2 A_{\mu}^{ab} \gam_{ab} dx^{\mu}$ where $\gamma_{ab} = 1/2 [\gamma_a , \gamma_b ]$.
The field strength 2-form is given by 
$F = F(A) \equiv d A + g A^2 $ 
where $g$ is
the gauge coupling constant.  Similarly, a covariant derivative 1-form
of $\phi$ is given by $D_A \phi \equiv d \phi + g [A , \phi]$.
Under a gauge transformation, 
$A \go \Omega A \Omega^{-1} + (1/g) \Omega d \Omega^{-1}$, 
$F \go \Omega F \Omega^{-1}$, and 
$D_A \phi \go \Omega D_A \phi \Omega^{-1}$,
where $\Omega = \exp \{ \varepsilon_{ab}(x) \gam^{ab}  \}$

The action is given by 
\beqn
I &\equiv& 
\int \left[ \frac{1}{8} \Tr F^2  * F^2 
+ \frac{1}{8} \Tr D_A \phi *D_A \phi 
- {\lambda}  (\phi^a \phi_a -H_0^2 )^2 d^6 x \right] \cr
\noalign{\kern 10pt}
&=& \int d^6 x \, 
\left\{ -\frac{1}{8 \cdot 4!} \Tr (F^2)_{\mu \nu \rho \sigma} 
  (F^2)^{\mu \nu \rho \sigma} 
  - \frac{1}{2}D_{\mu} \phi^a D^{\mu} \phi_a 
  - \lambda ( \phi^a \phi_a - H_0^2)^2 \right\} ~~.
  \label{action1}
\eeqn
Here the components of $F^2 = \frac{1}{8} \{F_{\mu \nu} , F_{\rho \sigma}  \}
  dx^\mu \wedge  dx^\nu \wedge dx^\rho \wedge dx^\sigma$ are given by 
\beqn
&&\hskip -1cm 
(F^2)_{\mu \nu \rho \sigma} 
= T_{\mu \nu \rho \sigma}^e \gamma_e  - S_{\mu \nu \rho \sigma} \cr
\noalign{\kern 10pt}
&&\hskip -1cm 
T_{\mu \nu \rho \sigma}^e(A) = 
\frac{1}{2 \cdot 4!}\eps^{abcde} 
\left( F_{\mu \nu}^{ab} F_{\rho \sigma}^{cd} 
  +F_{\mu \rho}^{ab} F_{\sigma \nu}^{cd} 
   + F_{\mu \sigma}^{ab} F_{\nu \rho}^{cd} \right) \cr
\noalign{\kern 10pt}
&&\hskip -1cm 
S_{\mu \nu \rho \sigma}(A) 
= \frac{1}{4!} \left( F_{\mu \nu}^{ab} F_{\rho \sigma}^{ab} 
  +F_{\mu \rho}^{ab} F_{\sigma \nu}^{ab} 
    + F_{\mu \sigma}^{ab} F_{\nu \rho}^{ab} \right) ~~,
\label{action2}
\eeqn
so that in the action   
$\frac{1}{4} \Tr (F^2)_{\mu \nu \rho \sigma}  (F^2)^{\mu \nu \rho \sigma}  
=  T_{\mu \nu \rho \sigma}^e  T^{\mu \nu \rho \sigma}_e 
+ 4 S_{\mu \nu \rho \sigma} S^{\mu \nu \rho \sigma}$.
The action of this type has been considered in ref.\ \cite{Tchrakian:1978sf}.  The relations in (\ref{action2}) 
are special to $SO(5)$ gauge theory. 

The action is quartic in $F_{\mu\nu}$, but is quadratic in $F_{0k}$.  
The Hamiltonian is positive semi-definite and is bounded from below 
by a topological charge.  To see it, first notice that
\beqn
&&\hskip -1cm
T^e_{0jkl} = F^{ab}_{0i}\,  M^{ab,e}_{i,jkl} 
~~~,~~~
 M^{ab,e}_{i,jkl} 
 = {1\over 2 \cdot  4!} \,\eps^{abcde} L^{cd}_{i,jkl} ~~~, \cr
\noalign{\kern 10pt}
&&\hskip -1cm
S_{0jkl} = F^{ab}_{0i}\,  N^{ab}_{i,jkl} 
~~~,~~~
 N^{ab}_{i,jkl} 
 = {1\over  4!} \,  L^{ab}_{i,jkl} ~~~, \cr
\noalign{\kern 10pt}
&&\hskip -1cm
L^{cd}_{i,jkl} = 
\del_{ij}F_{kl}^{cd}+\del_{ik}F_{lj}^{cd}+\del_{il}F_{jk}^{cd} ~~~.
\label{MNdef}
\eeqn
The canonical conjugate momentum fields are given by
\beqn
\Pi^{ab}_i &=& \frac{\del I}{\del \dot{A}_i^{ab}} 
=  \frac{1}{3!} T_{0jkl}^e 
   \frac{\del T_{0jkl}^e}{\del \dot{F}_{0i}^{ab}}
   +  \frac{4}{3!} S_{0jkl}  
     \frac{\del S_{0jkl}}{\del \dot{F}_{0i}^{ab}} \cr
\noalign{\kern 10pt}
&=& \frac{1}{3} ( M^{ab,e}_{i,jkl} M^{cd,e}_{m,jkl}
+ N^{ab}_{i,jkl}  N^{cd}_{m,jkl} ) F^{cd}_{0m} \cr
\noalign{\kern 10pt}
&\equiv& U^{ab,cd}_{i,m} F^{cd}_{0m} ~~.
\label{conjugate1}
\eeqn
$U$ is a symmetric, positive-definite matrix.
To confirm the positivity of the Hamiltonian, we take 
the $A_0=0$ gauge in which $F^{ab}_{0i} = \dot A^{ab}_i$.  It immediately follows that
\beeq
E = \int d^5x \, 
\left[ \frac{1}{2} \Pi U^{-1} \Pi 
+ \frac{1}{2 \cdot 4!} \Big\{ (T_{ijkl}^e)^2 
   + (S_{ijkl})^2 \Big\} + {\cal H}_{\phi} \right] \ge 0
\eneq
where ${\cal H}_{\phi}$ is the scalar field part of the Hamiltonian density.

In the $A_0=0$ gauge the energy becomes lowest for static configurations $\dot A_i^{ab} = \dot \phi_a= 0$.  It is given by
\beqn
E&=& 
\int d^5x \,  \myfrac{1}{4!} 
 \left[ \frac{1}{2} \, 
  (T_{ijkl}^e  \mp  \eps^{ijklm} D_m \phi^e )^2
    +\frac{1}{2} {S_{ijkl}}^2 
     \pm  \eps^{ijklm} T_{ijkl}^e D_m \phi^e
      + \lambda (\phi^a \phi_a -H_0^2 )^2 \right] \cr
\noalign{\kern 10pt}
&\geq& \pm \int d^5x \, 
  \frac{1}{4!} \eps_{ijklm} T_{ijkl}^e D_m \phi^e 
  = \pm \int \tr D_A   \phi \,F^2 \equiv \frac{16 \pi^2}{g^2} H_0 {\cal Q} ~~.
  \label{bound1}
\eeqn
As $D_A F=0$ and therefore $\tr D_A \phi \, F^2 = d(\tr \phi \, F^2)$,
${\cal Q}$ can be expressed as a surface integral
\be
{\cal Q} = \pm \frac{g^2}{16 \pi^2 H_0} \int_{S^4} \tr \phi \, F^2 ~~,
\label{charge1}
\ee
where $S^4$ is a space infinity  of $R^5$.

${\cal Q}$ is a charge $\int d^5x \, k^0$ of
a 6-dimensional current  $k^\mu$ defined by
$k = k_\mu dx^\mu = \pm * ({g^2}/{16 \pi^2 H_0}) \tr D_A \phi F^2$, which is 
 conserved, $d * k = 0$.  ${\cal Q}$ can also be viewed as a
 topological charge associated with Abelian Kalb-Ramond 
 3-form gauge fields whose 4-form field strength ${\cal G}$ is 
 given  by \cite{Tchrakian:1978sf} 
\bea
{\cal G} &=&  \tr \Big\{  
\hp \, F^2 + \frac{1}{2g} \hp\,  (D_A \hp)^2 F+
\frac{1}{16g^2} \hp (D_A \hp)^4 \Big\}  \cr
\noalign{\kern 5pt}
 &=&  \tr \hp \Big\{  F + \frac{1}{4g} (D_A \hp)^2 \Big\}^2 ~~.
 \label{Gfield1}
\eea
Here $\hp = \phi / | \phi |$, $| \phi | = \sqrt{\phi^a \phi^a}$
and $D_A \hp = d \hp + g[A, \hp]$.  


It is the salient feature of ${\cal G}$ given in (\ref{Gfield1})   that
it can be written as
\beqn
&&\hskip -1cm
{\cal G} = d {\cal C} + {1\over 16 g^2} ~ \tr \, \hp (d \hp)^4  ~~,\cr
\noalign{\kern 10pt}
&&\hskip -1cm
{\cal C} = {1\over 2g} ~ \tr \, \hp \, \bigg\{ (d \hp)^2 A 
 + g \, (d\hp \, A \, \hp \, A  + dA \, A + A \, dA)
   + g^2 \Big( A^3 + {1\over 3} \, A \, \hp \, A \, \hp \, A \Big) \bigg\} ~~.
\label{AbelianC}
\eeqn
${\cal C}$ does not have  a singularity of the Dirac string type where $|\phi| \neq 0$. 
${\cal G}$ and ${\cal C}$ are  the 't Hooft 4-form field strengths 
and the corresponding Kalb-Ramond 3-form  fields
in six dimensions, respectively.  
The expression (\ref{AbelianC}) is valid
in the entire six-dimensional spacetime.  We remark that the 
Kalb-Ramond 3-form fields ${\cal C}$ in (\ref{AbelianC}) is almost
the same as those in ref.\ \cite{Tchrakian2} where $A$ is replaced by
the asymptotic one which is valid only at $r \go \infty$ (on $S^4$).
  We also note that for configurations 
with $\hp = \gamma^5$, only gauge fields in the unbroken 
$SO(4)$,  
$\hA = \onehalf \sum_{a,b=1}^4  A_\mu^{ab} \gamma_{ab} dx^\mu$, 
contribute in (\ref{Gfield1}) and (\ref{AbelianC}).
Indeed, $\tr \, \gamma^5 (dA A + A dA) = \tr \, \gamma^5 (d\hA \hA + \hA d\hA)$
and $\tr \, [ \hp A^3 + \onethird (\hp A)^3 ] 
= {1\over 6} \tr \, \{ \hp, A \}^3 
= {1\over 6} \tr \, \{ \hp, \hA \}^3$.

As $D_A \hat \phi =0$ on $S^4$ at space infinity for any
configuration with a finite energy,  ${\cal G}$  coinsides
with $\tr \hat \phi \, F^2$ on $S^4$.  Hence
\beeq
{\cal Q} = \frac{g^2}{16 \pi^2 H_0} \int_{S^4} |\phi| {\cal G} 
= \frac{1}{256 \pi^2} \int_{S^4} \tr \, \hp (d \hp)^4 ~~.
\label{topological1}
\eneq
In the second equality we used the fact that ${\cal C}$ is regular in $S^4$ as $|\phi| \sim H_0$. 
The quantity appearing in the last equality in (\ref{topological1}) 
 is the winding number.  
The charge ${\cal Q}$ is thus regarded as the magnetic charge 
associated with Abelian Kalb-Ramond  field strengths ${\cal G}$. 

The Bogomolny bound equation is
\beeq
*_5 (F \wedge  F) = \pm D_A \phi
\label{BPS1}
\eneq
where 
 $*_5$ is  Hodge dual in five-dimensional space.
 In components it is given by 
\bea
\epsilon^{ijklm} T_{ijkl}^e &=& \pm D_m \phi^e ~~, \cr
S_{ijkl}&=&0 ~~.
\label{BPS2}
\eea
Let us define  $e\equiv x^a \gam_a / r$. We make a hedgehog 
ansatz\cite{Tchrakian2}
\beqn
\phi &=&  H_0 U(r) \, e ~~~, \cr
\noalign{\kern 10pt}
A &=& \myfrac{1-K(r)}{2g} \, e d e ~~~.
\label{ansatz1}
\eeqn
It follows immediately that
\beqn
D_A \phi &=& H_0( K U d e + U' e d r) ~~~, \cr
\noalign{\kern 10pt}
F &=& 
\frac{1-K^2}{4g} \,  d e \wedge d e 
  - \frac{K'}{2g} \, e d r \wedge d e ~~~.
  \label{ansatz2}
\eeqn
Boundary conditions are given by
$U(0)=0$, $K(0) = 1$, $U(\infty) = \pm 1$, and $K(\infty) = 0$.

With the use of 
$*_5 (de \wedge de \wedge de \wedge de ) = 4!\, e dr/ r^4$ and
$*_5 (e dr\wedge de \wedge de \wedge de ) = 3!\, de/r^2$,
the Bogomolny bound equation (\ref{BPS1}) (with a plus sign)
  becomes
\beqn
&&\hskip -1cm
 K U = - \frac{1}{\tau^2} ( 1- K^2 ) \myfrac{dK}{d\tau} ~~~,\cr
\noalign{\kern 10pt}
&&\hskip -1cm
 \myfrac{dU}{d\tau}
= \frac{1}{\tau^4}  ( 1 - K^2 )^2 ~~~, \cr
\noalign{\kern 10pt}
&&\hskip -1cm
\tau = a r ~~~, ~~~ 
a = \bigg( \frac{2 g^2}{3} H_0\bigg)^{1/3} ~~~.
\label{BPS3}
\eeqn
In this case $U$ increases  as $\tau$ so that $U(\infty) = 1$.
A solution in the case $- D_A \phi = *_5 (F \wedge F)$ is 
obtained  by replacing $U$ by $-U$.  We note that the two equations
in (\ref{BPS3}) can be combined to yield
\beeq
{d\over d\tau} \bigg( \myfrac{1-K^2}{\tau^2 K} {dK\over d\tau} \bigg) 
 + \myfrac{(1-K^2)^2}{\tau^4} = 0 ~~~,
 \label{BPS4}
 \eneq
or equivalently, in terms of $s=\ln \tau$ and $f(s) = K^2$,
\beeq
f'' - \bigg\{ 3 + \myfrac{f'}{f(1-f)} \bigg\} f' + 2 f (1-f) = 0 ~~.
\label{BPS5}
\eneq

The equation (\ref{BPS4}) with the boundary conditions 
$K(0) = 1$ and $K(\infty)=0$ is scale invariant, i.e.\
if $K(\tau)$ is a solution,  so is $K(\alpha \tau)$
with arbitrary $\alpha>0$. However, $U(\tau)$ changes, under this transformation,
to $\alpha^{-3} U(\alpha \tau)$ in (\ref{BPS3}) so that
the boundary condition $U(\infty)=1$ is fulfilled 
only with a unique value for $\alpha$.

The behavior of the solution near the origin is given by 
\bea
K &=& 1 - b \tau^2 + \frac{5}{14}b^2 \tau^4 + \cdots ~~~, \cr
\noalign{\kern 5pt}
U &=& 4 b^2 \tau \left\{1 - \frac{4}{7} b \tau^2  
  + \frac{20}{63} b^2 \tau^4 + \cdots   \right\}  ~~~.
  \label{expansion1}
\eea
The value of the parameter $b$ need to be determined such that
$U(\infty)=1$ is satisfied.  The behavior of the solution at
a space infinity $\tau=\infty$ is given by
\beqn
&&\hskip -1cm
K \sim K_0 e^{-\tau^3/3} ~~~, \cr
\noalign{\kern 5pt}
&&\hskip -1cm
U \sim 1 - {1\over 3\tau^3} ~~~.
\label{expansion2}
\eeqn
Note that $F \sim (4g)^{-1} de \wedge de$ and 
$D_A \phi \sim H_0 \tau^{-4} e d\tau$.

\begin{figure}[htb]
\centering 
\includegraphics[width=8.cm]{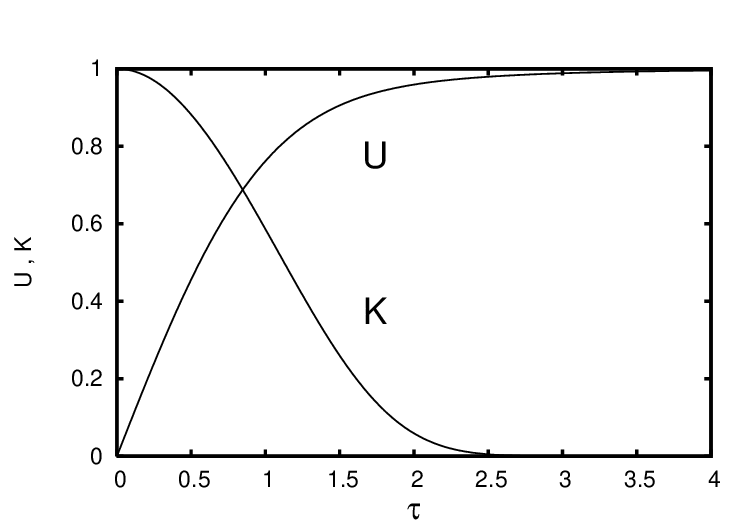} 
\caption{Solution : $U(\tau)$ and $K(\tau)$ in (\ref{ansatz1}).}
\label{UK1}
\end{figure}

A solution is obtained numerically.  We adopted the shooting method
to solve Eq.\ (\ref{BPS3}) from $\tau=0$ to $\tau=\infty$.
Precisely tuning the value of $b$ in (\ref{expansion1}), we find 
a solution with the boundary conditions $U(\infty)=1$ and 
$K(\infty)=0$.  It is  found that $b=0.494$ and $K_0 = 1.2$.  
The solution is displayed in fig.\ \ref{UK1}.

The energy, (\ref{bound1}), of the solution is given by
 $ \int \tr D_A \phi F^2$ in the 
$\lambda \go 0$ limit.  The insertion of (\ref{ansatz2}) leads, 
 with the aid of the identities 
 $\tr \, edr  (de)^4  = (4 \cdot 4!/r^4) d(volume)$ and 
  $\tr \, (de)^5 =0$,
to
\beqn
&&\hskip -1cm
E=   \int d^5 x \, {H_0 \over 16g^2}
 {4\cdot 4!\over r^4} \bigg\{ (1-K^2)^2 {dU\over dr}
  - 4 {dK\over dr} UK(1-K^2) \bigg\}  \cr
\noalign{\kern 10pt}
&&\hskip -.5cm
= {16\pi^2\over g^2} \, H_0 \Big[ (1-K^2)^2 U \Big]_0^\infty
= {16\pi^2\over g^2} \, H_0 ~~~.
  \label{energy2}
\eeqn
The same result follows from $E= (16 \pi^2 H_0 / g^2) {\cal Q}$ as ${\cal Q}= 1$.

As Dirac showed, a monopole configuration in $U(1)$ gauge theory 
in four dimensions necessarily has a Dirac string, or a singular
point in gauge potentials on the space infinity $S^2$.  
Quantization of Dirac strings, or monopole charges, corresponds to
nontrivial mapping around the singular point, or the hole,  
on $S^2$, namely $\pi_1 [U(1)]$.
The configuration of  a 't Hooft-Polyakov monopole in the 
$SO(3)$ gauge theory is regular everywhere and the monopole 
charge is related to the winding number of the Higgs fields 
which breaks $SO(3)$ to $U(1)$.  This fact is
summarized in the  exact sequence in the homotopy group
\beeq
\hbox{Ker} ~ \Big\{ \pi_1[U(1)] \go \pi_1[SO(3)] \Big\} 
 \simeq \pi_2(S^2) ~~. 
 \label{homotopy1}
 \eneq
 In our case $SO(5)$ gauge symmetry is broken to $SO(4)$ by the Higgs 
 fields $\phi^a$.  A monopole in $SO(4)$ gauge theory in six 
 dimensions accompanies a singularity in gauge potentials on  the
  space infinity $S^4$.  Quantization of monopole charges is 
  associated with $\pi_3[SO(4)]$.  The
 singularity is lifted by embedding $SO(4)$ into $SO(5)$, and 
 the monopole charge is reduced to the winding number $\pi_4(S^4)$ of 
 the Higgs fields.  The relation is summarized in 
 \beeq
\hbox{Ker} ~ \Big\{ \pi_3[SO(4)] \go \pi_3[SO(5)] \Big\} 
 \simeq \pi_4(S^4) ~~. 
 \label{homotopy2}
 \eneq
 Thus we observe that generalized monopoles in  $SO(5)$
 gauge theory in six dimensions described in the present paper
  are completely parallel to 
 't Hooft-Polyakov monopoles  in four dimensions.
 
 As another interesting aspect,  the generalized monopole solution  presented in this 
 paper may realize the electric-magnetic duality in the $M$-theory of strings.
 The 3-form Kalb-Ramond fields ${\cal C}$ defined in (\ref{AbelianC})
 couple to 2-branes in 11 dimensions. Dual of the field strength $d {\cal C}$
 is the 7-form field strength so that the  generalized monopole can be regarded as
 a source to the corresponding 6-form Kalb-Ramond fields, namely a 5-brane
 in 11 dimensional spacetime.  A similar argument applies to 2- and 4-branes in 
 ten dimensions.
 
In this paper we have shown that there exists a regular,
spherically symmetric monopole solution in the six-dimensional $SO(5)$ gauge 
theory with the action quartic in field strengths.  This is the first 
example of particle-like solitons in space dimensions bigger than four.
The energy in the $\lambda \go 0$ limit is given by the monopole
charge associated with the Abelian Kalb-Ramond 3-form fields.
The connection between $SO(5)$ gauge fields and the Kalb-Ramond 
fields is given by the generalized 't Hooft tensors ${\cal G}$
and ${\cal C}$. 
The solution is stable.
Physical consequences of  these generalized monopoles
are yet to be investigated. They  affect the evolution of
the universe at the very early stage, should there exist extra dimensions.
Their role for the compactification of extra dimensions and their relation
to extended objects in the matrix models need to be clarified.
Generalization of solutions to multi-monopole states is also 
awaited.  We hope to come back to these points in  future publications.

\vskip 1.cm

\leftline{\bf Acknowledgements}
This work was supported in part by  Scientific Grants from the Ministry of 
Education and Science, Grant No.\ 13135215 and
Grant No.\ 15340078 (Y.H.), and by Japan Society for the Promotion of Science 
under the Post-doctoral Research Program (M.N.).






\def\jnl#1#2#3#4{{#1}{\bf #2} (#4) #3}

\def\Zphys{{\em Z.\ Phys.} }
\def\jssc{{\em J.\ Solid State Chem.\ }}
\def\jpsJ{{\em J.\ Phys.\ Soc.\ Japan }}
\def\ptps{{\em Prog.\ Theoret.\ Phys.\ Suppl.\ }}
\def\PTP{{\em Prog.\ Theoret.\ Phys.\  }}

\def\JMP{{\em J. Math.\ Phys.} }
\def\NPB{{\em Nucl.\ Phys.} B}
\def\NP{{\em Nucl.\ Phys.} }
\def\PLB{{\em Phys.\ Lett.} B}
\def\PL{{\em Phys.\ Lett.} }
\def\PRL{\em Phys.\ Rev.\ Lett. }
\def\PRB{{\em Phys.\ Rev.} B}
\def\PRD{{\em Phys.\ Rev.} D}
\def\PRe{{\em Phys.\ Rep.} }
\def\AP{{\em Ann.\ Phys.\ (N.Y.)} }
\def\RMP{{\em Rev.\ Mod.\ Phys.} }
\def\ZPC{{\em Z.\ Phys.} C}
\def\SCI{\em Science}
\def\CMP{\em Comm.\ Math.\ Phys. }
\def\MPLA{{\em Mod.\ Phys.\ Lett.} A}
\def\IJMPA{{\em Int.\ J.\ Mod.\ Phys.} A}
\def\IJMPB{{\em Int.\ J.\ Mod.\ Phys.} B}
\def\EPJC{{\em Eur.\ Phys.\ J.} C}
\def\PR{{\em Phys.\ Rev.} }
\def\JHEP{{\em JHEP} }
\def\cmp{{\em Com.\ Math.\ Phys.}}
\def\JPA{{\em J.\  Phys.} A}
\def\JPG{{\em J.\  Phys.} G}
\def\NJP{{\em New.\ J.\  Phys.} }
\def\CQG{\em Class.\ Quant.\ Grav. }
\def\ATMP{{\em Adv.\ Theoret.\ Math.\ Phys.} }
\def\ibid{{\em ibid.} }

\renewenvironment{thebibliography}[1]
         {\begin{list}{[$\,$\arabic{enumi}$\,$]}  
         {\usecounter{enumi}\setlength{\parsep}{0pt}
          \setlength{\itemsep}{0pt}  \renewcommand{\baselinestretch}{1.2}
          \settowidth
         {\labelwidth}{#1 ~ ~}\sloppy}}{\end{list}}

\end{document}